\documentclass[aps,pra,showpacs,twoside,twocolumn,10pt]{revtex4-2}
\usepackage[colorlinks=true, citecolor=blue, urlcolor=blue, linkcolor = blue ]{hyperref}

\usepackage{amsthm,graphicx,bm,amsmath,xcolor,braket,plain,color,amsthm,amsfonts,tabularx,graphicx,bbm,mathtools,esvect,wrapfig,verbatim,enumitem,dcolumn,amssymb,appendix,physics,fmtcount,booktabs,csquotes,epsfig,times,geometry,array,mathrsfs,graphics,epstopdf,latexsym,comment,cancel}

\usepackage[normalem]{ulem}

\usepackage[mathscr]{euscript}
\def\Tr{\operatorname{Tr}}

\usepackage{hyphenat}

\usepackage{orcidlink}

\begin{document}

\title{
Metrological quantum-to-classical crossover in the volume of a noisy quasiperiodic lattice}

\author{Priya Ghosh\,\orcidlink{0009-0000-5908-9407}}
\email{priyaghosh1155@gmail.com}

\author{Debarupa Saha\,\orcidlink{0009-0006-3226-7799}} 
\email{debarupa73@gmail.com}

\author{Ujjwal Sen\,\orcidlink{0000-0002-0091-5847}} 
\email{ujjwal@hri.res.in}

\author{Debraj Rakshit\,\orcidlink{0000-0002-5227-0972}} 
\email{debrajrakshit@hri.res.in}

\affiliation{Harish-Chandra Research Institute, Chhatnag Road, Jhunsi, Prayagraj  211 019, India}
\affiliation{
Homi Bhabha National Institute, Training School Complex, Anushakti Nagar, Mumbai 400 094, India}

\begin{abstract}
Localization-delocalization transitions have recently been proposed for building a class of efficient quantum many-body critical sensors. 
This work scrutinizes metrological performances of such devices by focusing on the Aubry–André–Harper model that supports a localization-delocalization transition at finite strength of the onsite potential. We identify a metrological quantum-to-classical crossover driven by the interplay between noise and system size, whereby quantum-enhanced scaling of the quantum Fisher information persists only up to a finite, noise-dependent characteristic system-size.  We first consider thermal noise and show that, at and near the localization-delocalization transition, the quantum Fisher information exhibits quantum-enhanced scaling for small systems but the system is stripped of this 
advantage  
beyond the characteristic crossover length. The crossover length decreases with increasing temperature. We then consider imperfections in the lattice hopping strengths and find a qualitatively similar crossover. There, the quantum-enhanced regime, identified with super-extensive scaling, gives way to an extensive scaling--a classical-limited weaker form--at sufficiently large system sizes. Thus, distinct noise mechanisms lead to a common limitation on the scalability of quantum-enhanced sensing: increasing the probe size beyond a noise-dependent limit can destroy the metrological quantum advantage. 

\end{abstract}

\maketitle

\section{Introduction}
Near a quantum phase transition point in quantum many-body (QMB) systems, diverging length scale and enhanced susceptibility lead to strong responses to small parameter changes, which can be harnessed for quantum-enhanced parameter estimation~\cite{Helstrom1967,Helstrom1968,Helstrom1969,Caves1981,Holevo1982,Braunstein1994,Pezze2018,Ghosh2026}, giving rise to a new class of sensing devices, known as quantum critical sensors. QFI provides the ultimate bound on precision estimation of an unknown parameter~\cite{Helstrom1967,Helstrom1968,Helstrom1969,Caves1981,Holevo1982,Braunstein1994}. In a series of recent studies, it has been shown that QFI exhibits super-extensive scaling with system size  at quantum criticality~\cite{Boixo2007,Boixo2008,Mirkhalaf2020,Yang2022,He2023,Montenegro2024,Sahoo2025,Mondal2025,Cheng2025,Sarkar2025,Debnath2025,Debnath2026,  Agarwal2025,Yousefjani2025,Sahoo2026}. However, performance of these sensing devices strictly relies on several idealized scenarios, such as preparation of the system in the ground state in the thermodynamic limit, implicitly assuming arbitrarily large systems. It is, however, obvious that such idealistic constraints are not truly met within a realistic experimental set-up. Many-body quantum experiments are limited by finite size, finite temperature, and unavoidable  experimental challenges, such as fabrication defect.  As a result, it becomes necessary to scrutinize the scope of the proposed many-body metrological devices under these realistic constraints.

Sufficiently large deviations from the desired operating conditions gradually subdue quantum-enhanced scaling and strip the system of its metrological advantage, even though the underlying dynamics may still retain a certain degree of quantumness. This leads the system to enter a regime of operational classicality, in which small but finite quantum correlations may remain present at the microscopic level but are no longer operationally accessible for enhanced parameter estimation. Thus, it becomes crucial to understand the emergence of the operational classicality from a purely metrological point of view. Although enhanced sensitivity of the critical sensors is often reported in the thermodynamic limit, in practicality, experiments are necessarily performed on finite many-body systems. The crossover of metrologically useful regime with quantum advantage to the regime of operational classicality in the presence of noise thus demands a protocol- and platform-specific critical analysis in realistic, finite-sized systems, particularly in the criticality-based sensors.

In critical lattices, increasing the system size introduces a competition between two opposing effects. On the one hand, system size is a resource for ground state quantum sensing, since the quantum Fisher information, quantified via fidelity susceptibility, grows as $F_Q \sim L^{\alpha}$ with $\alpha > 1$~\cite{Montenegro2024,Agarwal2025,Ghosh2026}. On the other hand, the same increase in system size amplifies the impact of noise. The typical low-lying level spacing decreases with increasing system size. As a result, thermal fluctuations populate an extensive number of states and the encoded state departs from the ground state. This affects adversely on the performance of the quantum sensors. Moreover, the uncertainty in the control parameters, such as the hopping strength, can only be dealt with configurational averaging that in turn destroys the metrological quantum advantage beyond a crossover system size.

These two competing effects lead to the metrological quantum-to-classical crossover (MQCC). Beyond a characteristic crossover size, set by the noise strengths under consideration, the averaging effects dominates, causing quantum-enhanced
regime to fade away and driving the system through the MQCC into the operationally classical metrological regime. Crucially, this competition implies that classicality in the critical systems is not solely determined by noise, such as temperature or uncertainty in parameter, but by a finite volume of the system over which quantum correlations remain operationally relevant. Thus, the noise introduces a characteristic length scale beyond which quantum effects average out.

The performances of the critical sensors have been investigated in the presence of noise. In this work, we demonstrate that finite size plays a crucial role in the metrologically motivated quantum-to-classical crossover. We consider a lattice in presence of a quasiperiodic potential, the well known Aubry-Andr{e'}-Harper (AAH) model. The AAH model undergoes a delocalization-localization crossover at  a finite strength of the quasiperiodic potential. Previously, it has been shown that adiabatic sensors with Heisenberg scaling can be designed by exploiting the localization crossover in the ground state~\cite{Sahoo2024,Sahoo2026}. In this work we study the estimation of the onsite potential strength via the QFI of thermal states. We show that thermal noise does not reduce sensitivity uniformly, but instead, it induces a sharp crossover in the system size and becomes effectively operationally classical. This identifies a finite  volume of the system, within which quantum states remain metrologically useful.

Beyond thermal effects, we investigate the influence of static statistical fluctuations in Hamiltonian parameters. Realistic lattice implementations are not devoid of fabrication imperfections. We consider that individual hopping strength may vary in each realization. For a particular realization, hopping strength remains fixed during the sensing protocol. We model it as a random variable drawn from a normal distribution. We therefore evaluate the QFI separately for each realization and subsequently perform a quenched average to compute QFI. Here, the width of the probability distribution essentially quantifies the strength of the noise. We report similar understandings--the system enters operational classicality at a finite, noise-dependent system size.

The rest of the paper is organized as follows. In Sec.~\ref{sec:pre}, we introduce the necessary theoretical concepts employed throughout this work, including the quantum Fisher information and the AAH model. In Sec.~\ref{sec:result}, we present the main results of this work by investigating the MQCC in the system size induced separately by thermal fluctuations and lattice defects.  We summarize our findings and conclude in Sec.~\ref{sec:conclusion}. Finally, the Appendices provides complementary studies on observable Fisher information and pairwise entanglement.

\section{Prerequisites}
\label{sec:pre}
In this section, we briefly review the QFI, the central figure of merit in quantum metrology, and introduce the AAH model considered in this work.

\subsection{Metrological figures of merit}
Let us suppose that the parameter to be estimated in a metrological protocol is given by $V$. In the framework of single-parameter quantum metrology, the estimation error $\Delta V$ corresponding to the parameter of interest $V$ is bounded by the quantum Cramér–Rao bound (QCRB)~\cite{Helstrom1967,Helstrom1968,Helstrom1969,Caves1981,Holevo1982,Braunstein1994,Pezze2018,Ghosh2026} as
\begin{align*}
   \Delta V \geq \frac{1}{\sqrt{\mathbf{F}_\mathbf{Q}(V)}},
\end{align*}
where $\mathbf{F}_\mathbf{Q}(V)$ is the quantum Fisher information (QFI) associated with $V$.  The QFI for the parameter $V$ can be expressed as~\cite{Zanardi2008}
\begin{align}
 \mathbf{F}_\mathbf{Q}(V) = 4 \chi(V) = 8 \left(\frac{1 - \tilde{F}(V)}{(\delta V)^2}\right), \label{QFI-mathematical-expression}
\end{align}
here $\chi(V)$ is called the fidelity susceptibility, $\tilde{F}(V) \coloneqq \Tr \sqrt{\rho(V)^{1/2} \, \rho(V + \delta V) \, \rho(V)^{1/2}}$. And $\delta V \to 0$ represents an infinitesimal change in the parameter $V$. Furthermore, $\rho(V)$ denotes the quantum state encoding the parameter $V$, to be estimated. 

For single-parameter estimation, there always exists at least one measurement that saturates the QCRB. Throughout this work, we use Eq.~\eqref{QFI-mathematical-expression} to calculate QFI for any parameter of interest.  
The QFI generally scales with the system size $L$ as $\mathbf{F}_\mathbf{Q} \sim L^\alpha$, where $\alpha$ is called the scaling exponent. The case $\alpha = 1$ is known as the standard quantum limit (SQL), while $\alpha = 2$ is referred to as the Heisenberg limit. $\alpha > 1$ signify \emph{quantum-enhanced sensitivity}~\cite{Giovannetti2006,Giovannetti2011}.

\subsection{AAH model}
 We consider a one-dimensional fermionic lattice with an quasi-periodically modulated onsite potential. The Hamiltonian is given by~\cite{Aubry1980},
\begin{align}
    \hat{H} = - J \sum_{i=1}^{L} (\hat{c}_i^\dagger \hat{c}_{i+1} + h.c) + V \sum_{i=1}^L \cos (2 \pi \iota \omega ) \hat{c}_i^\dagger \hat{c}_i.
\end{align}
We consider the single-particle scenario under periodic boundary conditions, $L+1=1$. Here, $J$ and $V$ denote the hopping strength and the on-site potential strength, respectively.  $\omega$ is an irrational number. 

Such a Hamiltonian exhibits a localization-delocalization transition at $V_c=2J$. For $V<V_c$, the system is in the delocalized phase, where the energy eigenstates are extended over the lattice and the particle is spread across multiple lattice sites. In contrast, for $V>V_c$, the system enters the localized phase, where the energy eigenstates become spatially localized, with the particle predominantly confined within a finite region of the lattice.  This transition point is energy independent, implying that all single-particle eigenstates undergo the delocalization-to-localization transition at the same critical potential strength, $V_c/J=2$. Experimental observations of localization to delocalization crossover in the AAH model have been reported, e.g., in ultracold atom set-up~\cite{Schreiber2015,Negro2003,Roati2008,Modugno2010,uab} and in photonic crystals~\cite{Lahini2009,Kraus2012,Verbin2013,Verbin2015}. Throughout the remainder of this work, we set $J=1$ for convenience. 

As discussed in~\cite{Sahoo2024},  the appropriate finite-size scaling  behavior at the transition is obtained by choosing the system sizes from either the odd or the even sequence of the Fibonacci series. Accordingly, throughout this work, we choose the system size $L$ is chosen as a Fibonacci number, $L=\mathcal{F}_n$, and the irrational parameter $\omega$ is approximated by the ratio of two consecutive Fibonacci numbers, $\omega=\mathcal{F}_n/\mathcal{F}_{n+1}$. Here, $\mathcal{F}_n$ and $\mathcal{F}_{n+1}$ are consecutive Fibonacci numbers. In the thermodynamic limit, this ratio approaches
$\omega=\lim_{n\to\infty}\frac{\mathcal{F}_n}{\mathcal{F}_{n+1}}=\frac{\sqrt{5}-1}{2}$, which is the inverse golden ratio. 

Ref.~\cite{Sahoo2024} demonstrated that the ground-state QFI corresponding to the parameter of interest $V$ of the AAH model at $V_c = 2$ admits Heisenberg scaling, $\mathbf{F}_\mathbf{Q} \sim L^2$.  This protocol corresponds to the idealized noiseless estimation scenario of the on-site potential strength. 

This result motivates us to investigate whether such enhanced metrological scaling persists in the presence of noise during the encoding process. In particular, we consider two types of noise, namely thermal noise and lattice-defect noise. In the presence of such noise, the encoded state is generally no longer the ground state of the Hamiltonian. Therefore, we investigate how the QFI scales with the system size $L$ when the parameter $V$ is encoded in a noisy setting and the initial state is not restricted to the ground state. In particular, we ask whether an quantum advantage , $\mathbf{F}_\mathbf{Q}\sim L^{\alpha}$, with $\alpha > 1$,  may persist in the presence of such practical considerations.

\section{Noise-induced MQCC in the volume of the system}
\label{sec:result}
In this section, we consider two distinct noise scenarios. We first investigate the effect of thermal noise on the estimation of (V), and subsequently examine lattice defects as another experimentally relevant source of noise.

\subsection{Crossover induced by thermal noise}
\label{subsec-thermal}
Here, we investigate the effect of the temperature-induced thermal noise on estimation of the on-site potential strength, $V$, of the AAH model. In the noiseless scenario, corresponding to the zero-temperature limit, the probe state can be taken to be the ground state of the Hamiltonian. In the presence of thermal fluctuations at finite temperature, however, contributions from excited energy eigenstates comes into play, and the probe is instead described by the thermal state
\begin{equation}
\label{eq:thermal-state}
    \rho_\beta=\frac{e^{-\beta\hat H}}{Z}, \quad Z=\operatorname{Tr}(e^{-\beta\hat H}),
\end{equation}
where the inverse temperature, $\beta$, is defined as, $\beta=1/T$.  Considering $\rho_{\beta}$ as the encoded state, we analyze the corresponding QFI, $\mathbf{F}_\mathbf{Q}$, and 
investigate the metrological performance with varied system sizes.

At the critical point, $V_c=2$, the system is expected to exhibit maximal metrological sensitivity: for the finite system sizes investigated in Ref.~\cite{Sahoo2024}, the ground-state QFI was found numerically to attain its maximum at, or in the immediate vicinity of, the critical point $V_c=2$, relative to the other values of $V$ considered. This behavior persists across the numerically accessible range, including the largest system sizes studied, indicating effectiveness of the system size as a metrological resource in these proposed quantum critical sensing devices, albeit in an idealized scenario. Here we investigate how the QFI of the thermal-encoded state depends on the system size $L$ at different temperatures in the vicinity of the critical point $V_c=2$. We begin by analyzing the scaling of the QFI with L exactly at $V=V_c$, and subsequently examine how this behavior changes as the on-site potential strength is tuned slightly away from the critical value.

\begin{figure*}[t]
    \centering
    \includegraphics[scale=0.45]{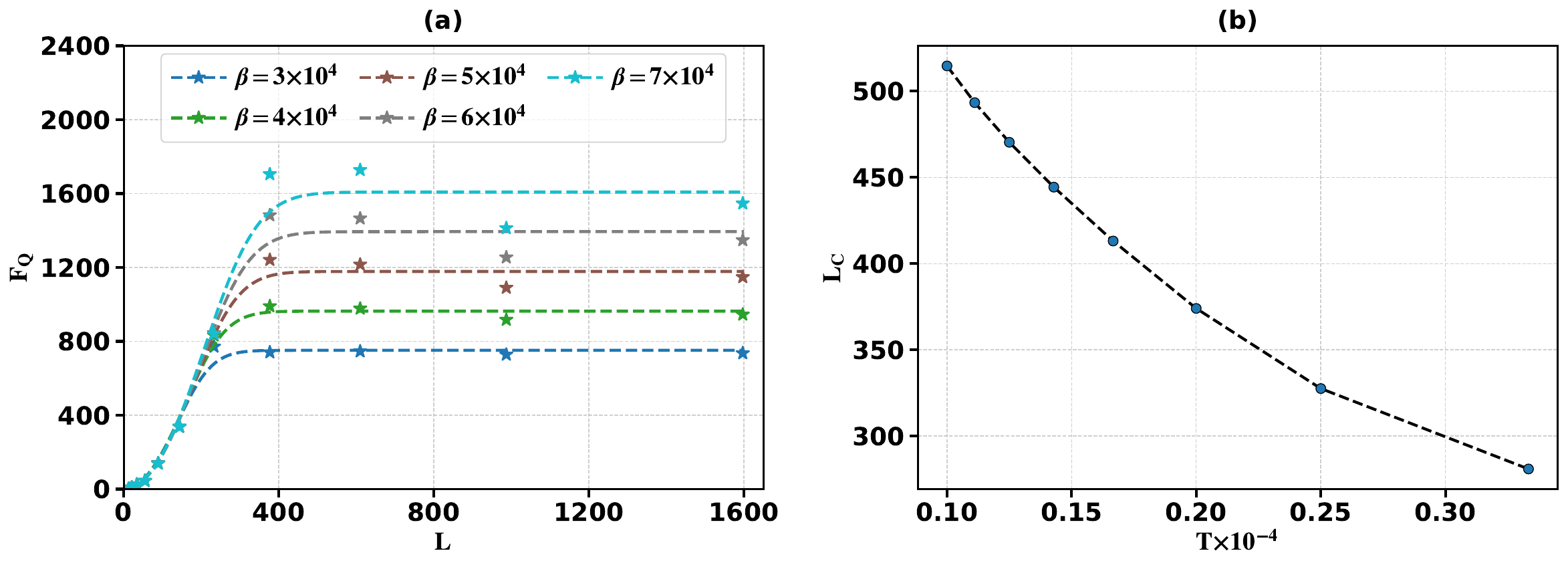}
    \caption{\textbf{Thermal noise-induced MQCC.} (a) shows QFI, $\mathbf{F}_\mathbf{Q}$, for the the on-site potential strength, $V$, corresponding to the thermal state (vertical axis) as a function of the system size, $L$ (horizontal axis).  For every $\beta$, the plot depicts the existence of a characteristic length $\mathbf{L}_\mathbf{C}(\beta)$, below which the QFI scales as $L^2$, and above which QFI nearly saturates, indicating the loss of quantum advantage. The plot thus establishes the system's departure from an quantum enhanced regime to entrance into an operationally classical regime through the MQCC with increasing system size. All the axes are dimensionless.
   (b) depicts the variation of $\mathbf{L}_\mathbf{C}$ with the temperature, $T$. As the temperature decreases, $\mathbf{L}_\mathbf{C}(\beta)$ shifts towards larger system sizes. This behavior is expected, as weaker thermal fluctuations allow the system to retain its quantum features over a larger volume, thereby extending the regime in which the enhanced QFI scaling persists. The vertical axis is dimensionless, whereas the horizontal axis is in units of $J/k_B$. 
   }
    \label{fig:thermal noise first}
\end{figure*}

We plot QFI as a function of system size $L$ for various inverse temperatures $\beta$ in Fig.~\ref{fig:thermal noise first}(a). Within the temperature window explored numerically in this work, the QFI consistently exhibits the same qualitative feature, namely, a thermal-noise-induced MQCC. This transition is manifested as a change in the scaling of the QFI with the system size:
\begin{itemize}
    \item
    For small system sizes $L$, the QFI exhibits Heisenberg scaling, $\mathbf{F}_\mathbf{Q}(V) \sim L^2$, indicating a quantum-enhanced estimation regime.
    \item 
    As the system size increases beyond a characteristic length scale, $\mathbf{L}_\mathbf{C}$,  the QFI saturates to a constant, $\mathbf{F}_\mathbf{Q}(V) \sim L^0$.
    Consequently, the quantum advantage does not persist for large-enough $L$, in contrast to the noiseless ground-state scenario. Instead, the estimation enters a operationally classical regime in which the QFI becomes effectively independent of the system size.
\end{itemize}
Thus, we have a noise-induced MQCC with respect to the system volume: the system exhibits quantum-enhanced scaling at small $L$, while the QFI becomes independent of $L$ in the large-volume regime, indicating the progressive loss of quantum advantage as the volume of the system increases.\\

This behavior can be understood from the increasing density of energy levels with system size. We characterize the relevant energy scale by the typical spacing between neighboring low-lying energy eigenstates, denoted by $\Delta E_{\mathrm{typ}}$. At a fixed inverse temperature $\beta$, increasing the system size $L$ leads to an increasingly dense energy spectrum and hence to a decrease in $\Delta E_{\mathrm{typ}}$. The thermal population of an excited state $|n\rangle$ relative to the ground state $|0\rangle$ is given by,
\begin{equation}
    \frac{p_n}{p_0}
    =
    \exp\left[-\beta(E_n-E_0)\right],
\end{equation}
where $E_n-E_0$ is the energy gap between the excited state and the ground state. When the characteristic energy scale of the low-lying excitations becomes comparable to or smaller than the thermal energy,  $k_{\mathrm{B}}T=1/\beta$, i.e.,
\begin{equation}
    {\beta}~\Delta E_{\mathrm{typ}}\lesssim {1},
\end{equation}
excited states acquire appreciable thermal populations. Thus, for a given $\beta$, there exists a characteristic system size $\mathbf{L}_\mathbf{C}(\beta)$ beyond which the increasing density of energy levels leads to significant thermal occupation of excited states. For $L\ll \mathbf{L}_\mathbf{C}(\beta)$, the low-lying energy levels are sufficiently well separated compared with the thermal energy, and the thermal state remains predominantly in the ground state. As $L$ approaches and exceeds $\mathbf{L}_\mathbf{C}(\beta)$, an increasing number of excited states become thermally accessible, resulting in stronger thermal mixing and a suppression of the quantum-enhanced QFI scaling.

The characteristic size $\mathbf{L}_\mathbf{C}(\beta)$ itself depends on the inverse temperature. As $\beta$ increases, corresponding to lowering the temperature, the thermal energy scale $1/\beta$ decreases. Consequently, a smaller energy spacing is required for excited states to acquire appreciable thermal populations. Since $\Delta E_{\mathrm{typ}}$ decreases with increasing $L$, the condition
\begin{equation}
    {\beta}~\Delta E_{\mathrm{typ}}(\mathbf{L}_\mathbf{C})\sim {1}
\end{equation}
is reached only at a larger system size for larger $\beta$. Hence, increasing $\beta$ shifts the crossover to larger $L$, extending the range of system sizes over which the thermal state remains predominantly ground-state-like and the quantum-enhanced metrological scaling persists.

Thus, $\mathbf{L}_\mathbf{C}(\beta)$ characterizes the crossover between the quantum-enhanced regime at small $L$ and the saturation regime at large $L$. To identify this characteristic length scale, we find that the curves in Fig.~\ref{fig:thermal noise first} are well described by the fitting function, 
\begin{equation}
    \mathbf{F}_\mathbf{Q}(V) \sim a\tanh(bL^2),
    \label{fitting}
\end{equation}
for different values of $\beta$.
We then proceed to determine $\mathbf{L}_\mathbf{C}(\beta)$ from the fitted function. Specifically, $\mathbf{L}_\mathbf{C}(\beta)$ denotes the value of $L$ at which the curvature of the QFI as a function of $L$ changes sign, i.e.,
\begin{equation}
    \frac{\partial^2 \mathbf{F}_\mathbf{Q}(V)}{\partial L^2}=0.
    \label{inflation}
\end{equation}
Figure~\ref{fig:thermal noise first}(b) shows the variation of $\mathbf{L}_\mathbf{C}(\beta)$ with temperature $T$. As $\beta=1/T$ increases, corresponding to a decrease in temperature, $\mathbf{L}_\mathbf{C}(\beta)$ shifts towards larger system sizes. Thus, the characteristic length scale increases with increasing $\beta$, indicating that the quantum-enhanced regime persists over a larger range of system sizes at lower temperatures.

Note that this entire analysis was done at $V_c=2$. To probe whether this behavior persists close to $V_c$, we extend our analysis to the system-size dependence of the QFI at different temperatures for several values of $V$ in the vicinity of the critical point.
\begin{figure*}[t!]
    \centering
    \includegraphics[scale=0.45]{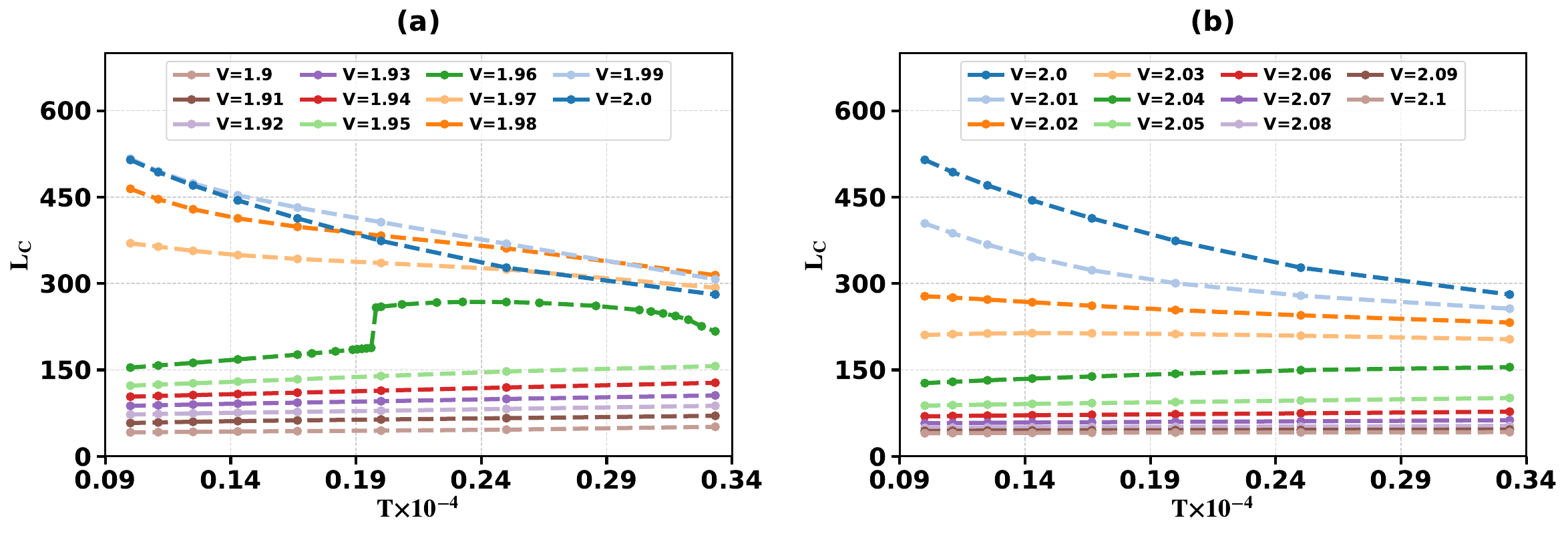}
    \caption{We plot $\mathbf{L}_\mathbf{C}$ as a function of temperature, $T$, for varied values of $V$ when (a) $\delta<0$ and when (b) when $\delta>0$. Note that the trend of $\mathbf{L}_\mathbf{C}$ decreasing with increasing $T$, obtained at $V_C = 2$, starts to fade away as $\delta$ increases. At sufficiently flat $\delta$, the trend becomes nearly flat.  The vertical axis is dimensionless, whereas the horizontal axis is in units of $J/k_B$.}
    \label{fig:thermal noise}
\end{figure*}

We numerically probe the nearby regime of $V_c=2$ and find the to remain overall picture remains qualitatively consistent throughout. For different  examined choices of $V$, such that $(V_c-V)=\delta$, with $|\delta| \ll V_c$, the QFI exhibits the similar characteristic system-size dependence as shown in Fig.~\ref{fig:thermal noise first}(a). Specifically, at a fixed temperature, the QFI initially displays the quantum-enhanced scaling $\mathbf{F}_\mathbf{Q}(V)\sim L^2$, beyond which the thermal fluctuations become dominant, causing the QFI to saturate, $\mathbf{F}_\mathbf{Q}(V)\sim L^0$. This signifies a thermal-noise-induced MQCC in the system, at or near the quantum critical point. However, for a given $\beta$, $\mathbf{L}_\mathbf{C}(\beta)$ changes with $V$.  We find the corresponding $\mathbf{L}_\mathbf{C}(\beta)$ for different temperatures and on-site potentials around the critical point via a similar fitting technique discussed in context of $V_c=2$ (see Eqs.~(\ref{fitting}-\ref{inflation}).

Figure \ref{fig:thermal noise} shows this finite-size analysis of $\mathbf{F}_\mathbf{Q}$ for varied choices of $V$ as a function of $T$, where the left (right) panel corresponds to $V<V_c$ ($V>V_c$). The plots unveil an interesting feature--the feature of temperature-induced MQCC appearing at a crossover system size fades as the system is tuned away from the critical point on either side. While close to $V_c$, $\mathbf{L}_\mathbf{C}(\beta)$ decreases significantly with increasing temperature; this temperature dependence gradually diminishes with increasing $|\delta|$. Sufficiently far from criticality, $\mathbf{F}_\mathbf{Q}$ remains devoid of any quantum advantage throughout the system-size and temperature ranges, including the ground state. Consequently, away from $V_c=2$, the temperature naturally has little to no scope for impact, since the quantum advantage vanishes in the zero-temperature limit itself. Equivalently, as the on-site potential moves farther from $V_c$, $\mathbf{L}_\mathbf{C}$ approaches a nearly constant value. This indicates a strong interplay between critical quantum effects and thermal fluctuations as a distinctive feature of the critical region.

We provide a complementary analysis examining whether the thermal-noise-induced MQCC identified through the QFI is also reflected in an experimentally accessible metrological quantity in  Appendix A. There, we study the observable Fisher information associated with a suitable experimentally relevant operator at the critical point. We find that the observable Fisher information captures the same qualitative features associated with MQCC

In a quantum critical sensor, the enhanced sensitivity generally emerges from the combined influence of several critical features, making it difficult to isolate the specific contribution of quantum correlations. It is nevertheless interesting to examine their fate independently in the presence of noise. Appendix B provides such a parallel study by investigating the pairwise entanglement content of the thermal state, quantified via logarithmic negativity.

In the subsequent section we analyze the effect on the scaling of QFI, considering yet another type of noise that is manifested in the form of lattice defects.

\subsection{MQCC induced by lattice defect}
\label{subsec-lattice-defects}
So far, our investigation has been considered noise in the form of thermal fluctuations. However, there can be other sources of practical situations, e.g., lattice defect. Within the conventional single-parameter estimation protocol, the other (control) parameters are assumed to be precisely known. In realistic experimental implementations of the AAH model, fabrication imperfections can cause individual hopping amplitudes to deviate from their desired values, and such disordered hopping configurations have been considered in several previous studies~\cite{Sherrington1975,Edwards1975,Derrida1980,Sadhukhan2015Beating, Bera2017Spontaneous,Mishra2016Constructive,Sadhukhan2016,Bera2019,Das2019,Bhattacharyya2024, Vishnupriya2026,Ghosh2026b}. Sensing can be susceptible to such experimental preparation imperfections of the control parameters in the Hamiltonian. In this context, a recent study has addressed the issue of parameter uncertainty in a second-order phase transition-based quantum critical sensor~\cite{Mihailescu2025Uncertain}, the known parameters, are considered to be limited by experimental resolution  and are shown to be potentially affect the performance of the sensing devices. 

Motivated by the consideration of lattice defect due to fabrication imperfection, in this subsection, we investigate the effect of disorder in the hopping amplitude of the AAH model, $J$, while retaining the onsite potential, $V$, as the parameter to be estimated. We model this uncertainty by drawing the hopping amplitudes, $J_i$, independently from a Gaussian distribution with mean unity and a sufficiently small standard deviation, $\sigma$. The small standard deviation ensures that the introduced disorder corresponds to weak lattice imperfections while preserving the average hopping strength of the ideal system. With this model disorder, we fix the on-site potential strength at the localization-delocalization crossover point, $V=V_c=2$. 

\begin{figure*}
    \centering
    \includegraphics[scale=0.45]{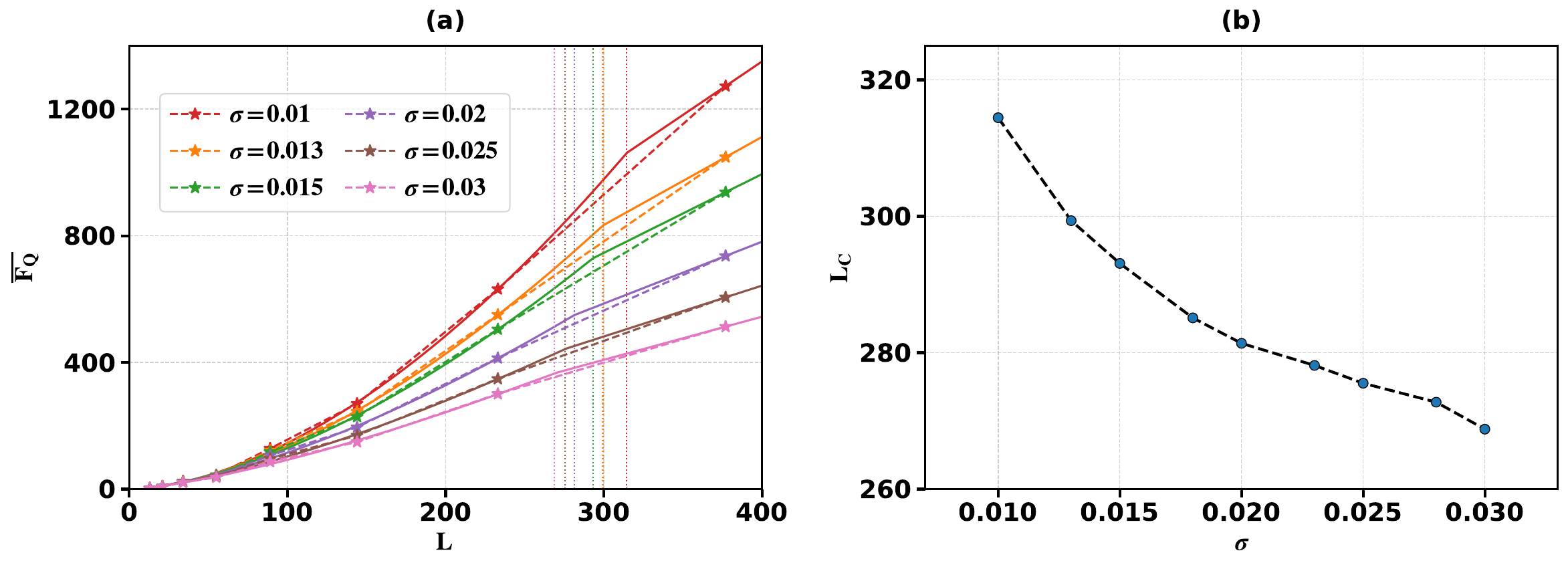}
    \caption{\textbf{Lattice-defect-induced QMCC.} (a) depicts quenched averaged ground-state QFI, $\overline{\mathbf{F}_\mathbf{Q}}$,  corresponding to the ground state of the AAH model in the estimation of the on-site potential strength of the AAH model (vertical axis) with system size (horizontal axis) for different choices of $\sigma$. The plot demonstrates $\sigma$-induced MQCC in the volume of the system. All the axes are dimensionless. (b) illustrates the crossover point, ($\mathbf{L}_\mathbf{C}$), with respect to the standard deviation of the Gaussian distribution, $\sigma$. As expected, the crossover point shifts towards high system size as the standard deviation of the Gaussian distribution decreases. All the axes are dimensionless.}
    \label{fig:external noise}
\end{figure*}
For each realization of the random hopping strength, we evaluate the corresponding QFI and subsequently perform quench averaging over a large number of random realizations in order to obtain the \emph{quench averaged} QFI. We then investigate whether the quantum-enhanced scaling persists in the presence of this uncertainty and determine how the MQCC depends on $\sigma$.

Considering ${\bf{J}}=(J_1,J_2, \cdots J_L)$ denotes a realization of the
hopping amplitudes, where $J_i$'s are drawn identically and independently from the Gaussian distribution
\begin{equation}
p(J_i)
=\frac{1}{\sqrt{2\pi}\sigma}
\exp\left[
-\frac{(J_i-J_0)^2}{2\sigma^2}
\right],
\end{equation}
where $J_0=1$ and $\sigma$ quantifies the strength of the lattice imperfections. Treating the hopping uncertainty as quenched disorder, the averaged over QFI, $\overline{\mathbf{F}_\mathbf{Q}}(V)$, is defined as 
\begin{equation}
    \overline{\mathbf{F}_\mathbf{Q}}\approx
\frac{1}{N}
\sum_{i=1}^{N}
\mathbf{F}_\mathbf{Q}\left[\rho({\bf{J}}^{(i)})\right],
\end{equation}
where $N$ denotes the number of realizations, and $\rho({\bf{J}}^{(i)})$ is the ground state density matrix corresponding to a particular $i^{\rm th}$ realization. 

We note that our treatment differs from that of Ref.~\cite{Mihailescu2025Uncertain}, where the QFI is evaluated for the uncertainty-averaged state. Here, we instead consider the lattice defects to be quenched, so that the hopping configuration remains fixed within each realization~\cite{Bera2014Classical, Bera2016Disorder}. Correspondingly, QFI is evaluated for each realization. A configurational averaging over many realizations is performed in the end. Our approach thus characterizes the typical scenario of deterministic evaluation of the metrological performance in each realization. In this analysis, we restrict our interest to the ground state only.

For the ideal AAH model with $J=1$, the localization-delocalization crossover at $V=V_c=2$ is expected to provide the highest metrological sensitivity. The introduction of lattice defects, however, tends to suppress this metrological enhancement. Consequently, the metrological behavior of the system is governed by the competition between the quantum enhancement originating from criticality and the degradation caused by the lattice defects.

Figure~\ref{fig:external noise}(a) shows the quenched QFI as a function of the system size $L$ for different disorder strengths. Similar to the thermal-noise case, we observe a lattice-defect-induced quantum-to-classical crossover. For small system sizes, the effect of the lattice defects is relatively weak, allowing the critical quantum effects to dominate. As a result, the quenched QFI exhibits the quantum-enhanced scaling
\[
\overline{\mathbf{F}_\mathbf{Q}}(V)\sim L^2.
\]
As the system size increases, however, the influence of the lattice defects on the many-body state becomes progressively more significant, gradually suppressing the critical quantum correlations responsible for the enhanced scaling. Beyond a characteristic crossover system size $\mathbf{L}_\mathbf{C}(\sigma)$, the quenched QFI no longer follows the quadratic scaling and instead follows
\[
\overline{\mathbf{F}_\mathbf{Q}}(V)\sim L,
\]
indicating that the quantum enhancement has been lost exhibits classical feature in the QFI scaling.

The numerical results are accurately captured by the piecewise fitting functions
\[
\overline{\mathbf{F}_\mathbf{Q}}(V) \sim aL^b,\qquad L\leq \mathbf{L}_\mathbf{C},
\]
and
\[
\overline{\mathbf{F}_\mathbf{Q}}(V)\sim a(\mathbf{L}_\mathbf{C})^{\,b-1}L,\qquad L>\mathbf{L}_\mathbf{C},
\]
from which the critical system size $\mathbf{L}_\mathbf{C}(\sigma)$ is calculated for each disorder strength.

The extracted values of $\mathbf{L}_\mathbf{C}$ for different disorder strengths are presented in Fig.~\ref{fig:external noise}(b). We observe that the critical system size increases systematically as the disorder strength decreases. This behavior indicates that weaker lattice defects are less effective in destroying the critical quantum correlations, allowing the quantum-enhanced ($L^2$) scaling to persist over larger system sizes. Increasing the disorder strength causes the MQCC to set in at progressively smaller values of $L$, demonstrating that stronger lattice imperfections suppress the critical metrological advantage rapidly. Consequently, the regime over which quantum-enhanced scaling can be observed shrinks monotonically with increasing lattice disorder.

\begin{figure*}
    \centering
    \includegraphics[scale=0.45]{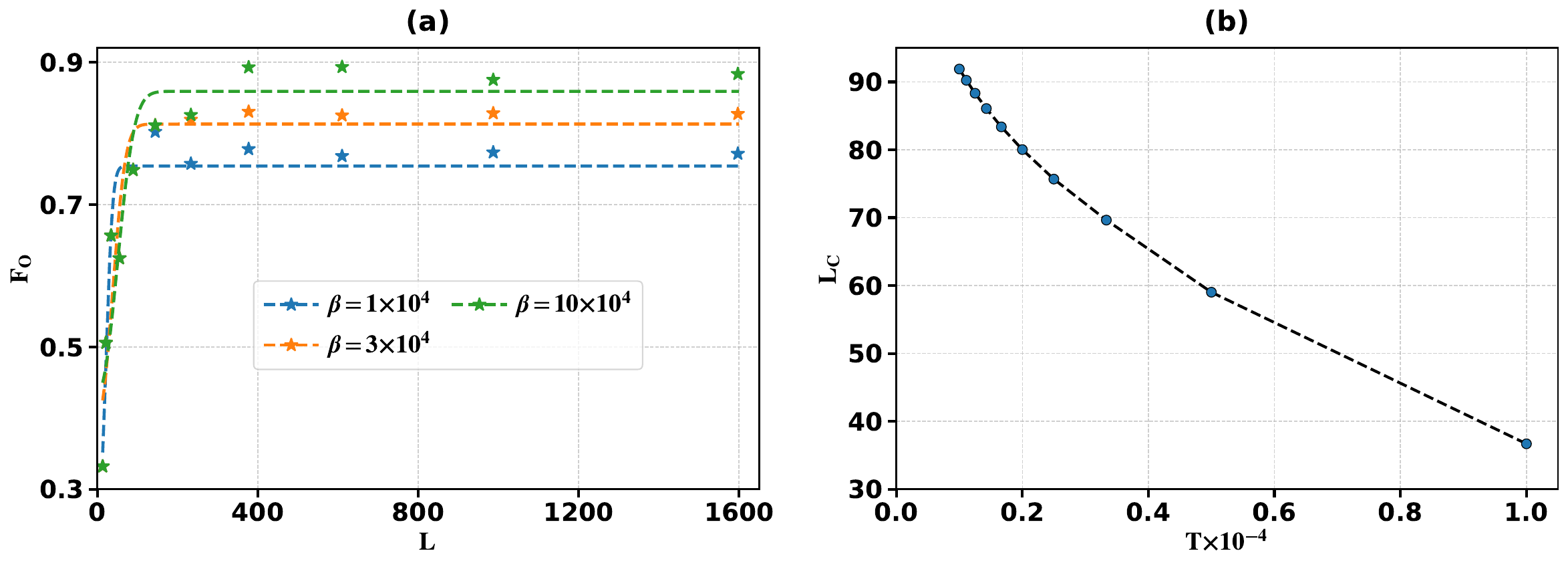}
    \caption{\textbf{Thermal-noise-induced MQCC reflected by OFI.} (a) We plot the OFI, for the thermal state of the AAH model in the estimation of the on-site potential strength $V$, as a function of the system size. Similar to the QFI, the OFI exhibits a thermal-noise-induced MQCC: the OFI initially displays quantum-enhanced scaling and subsequently saturates beyond a characteristic system size. All the axes are dimensionless. (b) The corresponding crossover system size ($\mathbf{L}_\mathbf{C}$) is plotted as a function of temperature. Consistent with the QFI analysis, the crossover shifts towards larger system sizes as the temperature decreases. This behavior indicates that weaker thermal fluctuations allow the system to retain its quantum correlations over larger system sizes, thereby extending the regime of quantum-enhanced metrological scaling reflected by the OFI. The vertical axis is dimensionless, whereas the horizontal axis is in units of $J/k_B$.
    }
    \label{fig:OFI}
\end{figure*}

\section{Conclusion}
In this work, we investigated how noise, namely thermal fluctuations and lattice defects, affect the metrological performances of quantum critical sensing devices. This question is particularly relevant in the context of quantum many-body systems, where noise and imperfections are unavoidable in realistic experimental implementations. In particular, we demonstrated that the  interplay of quantum effects, noise and system size gives rise to a MQCC, separating a quantum-enhanced metrological regime from an operationally classical one. We establish this by investing the AAH model, which supports localization-delocalization transition at a finite strength of the onsite potential.

We first considered thermal noise by taking the thermal state of the AAH Hamiltonian as the encoded probe. At finite temperature, thermal fluctuations populate excited energy levels, causing the probe to deviate from the ground state. We investigated the metrological behavior at the thermodynamic critical point, $V_c=2$, which separates the localized and delocalized phases of the AAH model, and as well as away from the critical regions. We found that the thermal-noise-induced MQCC in the QFI scaling persists not only at the critical point but also in its immediate vicinity. At sufficiently small system sizes, the Heisenberg scaling and hence the quantum advantage observed in the noiseless ground-state scenario are retained. However, beyond a characteristic system size, thermal fluctuations suppress this enhancement, and the QFI becomes independent of the system size, marking the system's entrance into an operationally classical metrological regime. The characteristic length scale depends on the temperature near the critical point, shifting towards larger system sizes as the temperature is lowered. As one moves farther away from the critical point, this characteristic length scale becomes essentially independent of the noise strength, indicating that the quantum-enhanced scaling is absent irrespective of the temperature. Thus, the noise-induced MQCC is intimately connected to the critical regime of the AAH model that supports quantum-enhanced sensing in the ideal (noiseless) limit and occurs as a function of the system volume. 

We further showed that MQCC is not specific to thermal noise. We considered lattice defects arising from the fabrication imperfection in the hopping strengths and computed the quenched average of the QFI over different disorder realizations. A qualitatively similar lattice-defect-induced MQCC is found in this case too. The QFI retains its quantum-enhanced scaling below a characteristic system size, while beyond this scale the enhancement is lost and the QFI exhibits a weaker, linear dependence on system size.

A common and robust feature of the two noise mechanisms is that the characteristic system size decreases as the noise strength increases. Increasing the temperature in the thermal case, or increasing the strength of the hopping disorder in the lattice-defect case, therefore progressively reduces the volume over which quantum-enhanced metrology can be sustained.

Finally, we note that a crossover in the system-size scaling of the QFI is not unique to noisy settings and may also occur when information is encoded in the ground state of a clean system. For example, Ref.~\cite{Mondal2025} reported that the ground-state QFI $F_{\mu\mu}$, associated with the estimation of the on-site potential $\mu$ in the one-dimensional Kitaev model, crosses over from $L^6$ scaling at smaller system sizes to $L^2$ scaling beyond a characteristic size. Importantly, both scaling regimes remain quantum enhanced, and hence this behavior represents a crossover between two quantum metrological regimes rather than an MQCC. By contrast, no analogous system-size-induced crossover was observed in the finite-size analysis of the clean AAH ground state in Ref.~\cite{Sahoo2024}, even for the largest system sizes accessible numerically. Therefore, within the AAH setting and the range of system sizes investigated here, the MQCC from quantum-enhanced to non-enhanced scaling can be attributed to the presence of noise. This finite-size behavior is directly relevant to engineered platforms, such as ultracold gases in optical lattices, where sensing protocols are necessarily implemented using finite systems, typically involving hundreds to thousands of lattice sites.

In summary, our results highlight an important aspect of the scalability of quantum-enhanced metrology. In realistic settings, increasing the size of a quantum probe does not necessarily lead to a persistent improvement in metrological performance. Instead, unavoidable noise can impose a finite, noise-dependent system size beyond which the quantum advantage is lost. The emergence of such a noise-induced crossover in system volume provides a physically relevant limitation on the scalability of quantum-enhanced sensing and suggests that, in the presence of realistic imperfections, identifying an appropriate system size may be as important as increasing the size of the probe. More broadly, our results demonstrate that the interplay between noise and system size can fundamentally determine the regime in which quantum resources remain useful for precision measurements.

\label{sec:conclusion}

\section*{Acknowledgment}
DS acknowledges support from the ‘INFOSYS scholarship for senior students’ at Harish-Chandra Research Institute, India. US acknowledges financial support from the Anusandhan National Research Foundation (ANRF), Government of India, under the Grant No. ANRF/ARG/2025/004617/PS.

\appendix

\section{Observable Fisher Information}
\label{sec:appendix}
To gain a broader understanding of the impact of thermal noise, here we analyze the effect of thermal noise on the system from two perspectives beyond QFI: (i) the observable Fisher information (OFI), which acts as another metrological figure of merit, and (ii)pair-wise, quantified by logarithmic negativity.

(i) \textit{ OFI:} In addition to the QFI, the precision of estimating the on-site potential strength $V$ can also be quantified through measurements of a suitable observable, known as observable Fisher information (OFI). In the asymptotic limit, the corresponding OFI associated with an observable $\hat{O}$ is obtained from the error-propagation formula, which is determined by the inverse of the signal-to-noise ratio (SNR)~\cite{Rams2018,Pezze2019},
\begin{equation}
\label{eq:error}
\mathbf{F}_\mathbf{O}^{-1}(V)
=
\lim_{\delta V\rightarrow0}
\frac{\langle \hat{O}^2\rangle-\langle\hat{O}\rangle^2}
{\left(\dfrac{d\langle\hat{O}\rangle}{d(\delta V)}\right)^2}.
\end{equation}
The OFI depends on the choice of the measured observable and therefore, in general, satisfies the inequality
\[
\mathbf{F}_\mathbf{O}(V,\hat{O})\leq \mathbf{F}_\mathbf{Q}(V),
\]
where the upper bound is given by the QFI. Upon optimizing over all possible observables, the maximum achievable OFI coincides with the QFI, thereby saturating the quantum Cram\'er-Rao bound~\cite{Cramer1946,Helstrom1969,Pezze2019}.

To examine whether the thermal-noise-induced metrological behavior observed through the QFI can also be captured by experimentally accessible observables, we consider
\[
\hat{O}=\sum_i\cos(2\pi i\omega)c_i^\dagger c_i,
\]
and evaluate the corresponding OFI for estimating the on-site potential strength $V$ at the critical point $V=V_c=2$, taking the thermal state of the AAH model as the encoded state.

The left panel of Fig.~\ref{fig:OFI} shows the OFI as a function of the system size $L$ for different temperatures. Remarkably, the OFI exhibits the same qualitative behavior as the QFI, demonstrating that the thermal-noise-induced MQCC is also reflected in observable-based metrology. For small system sizes, thermal fluctuations are sufficiently weak compared to the critical quantum correlations, and the OFI exhibits the quantum-enhanced scaling $\mathbf{F}_\mathbf{O}(V)\sim L^2$.
As the system size increases, however, thermal fluctuations progressively suppress the critical quantum correlations, leading to the saturation of the OFI, $\mathbf{F}_\mathbf{O}(V)\sim L^0$,
which corresponds to the classical metrological regime. The numerical results are well fitted by the fitting function
\[
\mathbf{F}_\mathbf{O}(V)\sim a\tanh(bL^2)+c,
\]
which captures the smooth crossover between the quantum-enhanced and classical scaling regimes.

To quantify this crossover, we extract the corresponding crossover system size from the OFI-versus-$L$ curves for each temperature, as shown in the right panel of Fig.~\ref{fig:OFI}. Similar to the behavior observed for the QFI, the crossover shifts towards larger system sizes as the temperature decreases. This indicates that weaker thermal fluctuations allow the critical quantum correlations to persist over larger systems, thereby extending the regime in which the OFI exhibits quantum-enhanced ($L^2$) scaling. The close agreement between the OFI and QFI results demonstrates that the thermal-noise-induced MQCC is not merely a feature of the ultimate precision bound but can also be detected through measurements of physically relevant observables.

\begin{figure}
    \centering
    \includegraphics[scale=0.35]{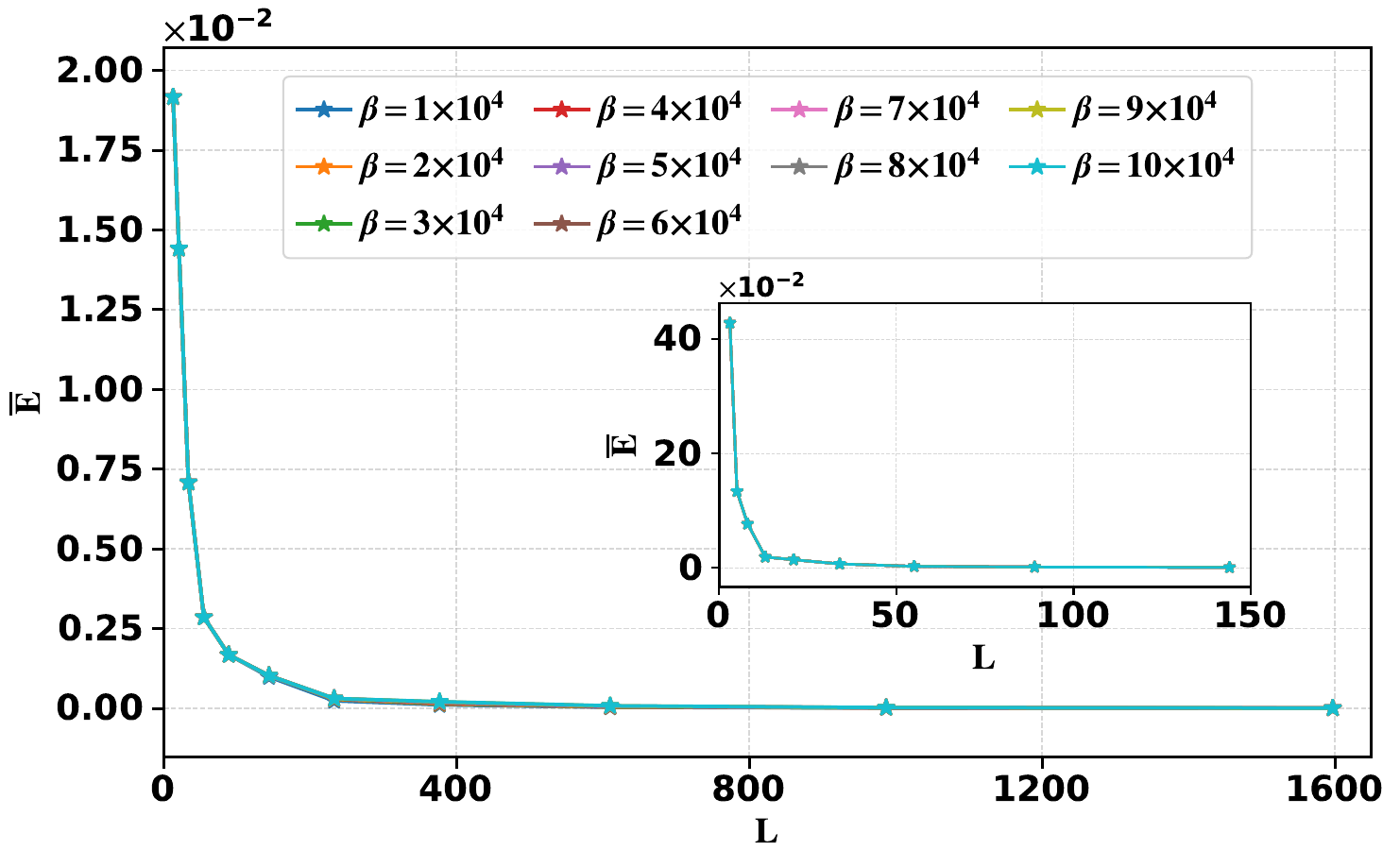} 
\caption{
\textbf{Average nearest-neighbor entanglement in the thermal state of the AAH model.}
The vertical axis shows the average logarithmic negativity, \(\overline{E}\), obtained by calculating the logarithmic negativity of every nearest-neighbor two-site reduced state, including the periodic bond \((L,1)\), and subsequently averaging over all \(L\) bonds. The horizontal axis denotes the system size \(L\). Results are presented for different temperatures, 
with the remaining model parameters chosen as in Sec.~\ref{subsec-thermal}. The main panel covers \(13\leq L\leq1597\), while the inset shows the behavior at smaller system sizes, \(3\leq L\leq144\). The average logarithmic negativity decreases rapidly with increasing \(L\), and the curves for different temperatures nearly overlap. Unlike the QFI and OFI, \(\overline{E}\) exhibits no discernible thermal-noise-induced crossover. Thus, the metrological crossover is not directly reflected in the average short-range bipartite entanglement of the multipartite thermal state.  The vertical axis is in ebits, while the horizontal one is dimensionless.
}
\label{fig:log-neg}
\end{figure}

\section{Entanglement} 
To investigate whether the thermal-noise-induced quantum-to-classical crossover observed through the QFI and OFI is accompanied by a corresponding change in the entanglement structure of the system, we analyze the nearest-neighbor bipartite entanglement present in the thermal state of the AAH model. For this purpose, we employ the logarithmic negativity~\cite{Negat2, Negat1,neg1,Negat3,Negat4,neg4}, which is well defined for bipartite mixed states.
Let
\begin{equation*}
\rho_{i,i+1} \coloneqq
\operatorname{Tr}_{L/\{i,i+1\}}
\left(\rho_{\beta}\right)
\end{equation*}
denote the reduced density matrix on the two-site subsystem $\{i,i+1\}$ of the thermal state $\rho_{\beta}$, given in Eq.~\eqref{eq:thermal-state}. The logarithmic negativity of this two-site state is defined as
\begin{equation*}
E_{\mathcal N}\left(\rho_{i,i+1}\right)
=
\log_{2}
\left\|
\rho_{i,i+1}^{T_i}
\right\|_{1},
\end{equation*}
where \(T_i\) denotes the partial transpose with respect to site \(i\), and \(\|\cdot\|_{1}\) is the trace norm.

To characterize the entanglement distributed over the lattice, we calculate the logarithmic negativity for every nearest-neighbor pair and then average over all bonds. Under periodic boundary conditions, the resulting average logarithmic negativity is
\begin{equation}
\overline{E}
=
\frac{1}{L}
\left[
\sum_{i=1}^{L-1}
E_{\mathcal N}\left(\rho_{i,i+1}\right)
+
E_{\mathcal N}\left(\rho_{L,1}\right)
\right].
\label{eq-avg-log-neg}
\end{equation}
Thus, \(\overline{E}\) measures the average short-range bipartite entanglement carried by the nearest-neighbor bonds of the multipartite system. It should, however, not be interpreted as a measure of genuine multipartite entanglement. 

We compute \(\overline{E}\) for different temperatures, using the same model parameters as in Sec.~\ref{subsec-thermal}. The results are presented in Fig.~\ref{fig:log-neg}, with the inset resolving the behavior for small system sizes. The average logarithmic negativity is appreciable for small \(L\), but decreases rapidly as the system size increases and becomes negligibly small for sufficiently large systems. Moreover, the curves corresponding to different inverse temperatures almost coincide over the considered temperature range, indicating that \(\overline{E}\) depends only weakly on temperature in this regime. It implies that \(\overline{E}\) does not display a temperature-dependent crossover with increasing system size analogous to that observed in the QFI and OFI. Therefore, the thermal-noise-induced metrological crossover is not mirrored by a corresponding crossover in the average nearest-neighbor bipartite entanglement. This demonstrates that the observed behavior of the QFI and OFI cannot be explained solely through the loss of short-range entanglement. Nevertheless, since \(\overline{E}\) does not capture genuine multipartite or long-range entanglement, the present result does not exclude the possible relevance of such correlations to the full many-body state.

\bibliography{ref}

\end{document}